\documentclass[reprint,floatfix,amsmath,amssymb,aps,prl]{revtex4-2}

\usepackage[usenames,dvipsnames]{xcolor}
\usepackage{graphicx}% Include figure files
\usepackage{dcolumn}% Align table columns on decimal point
\usepackage{bm}% bold math
\usepackage{hyperref}% add hypertext capabilities
\begin{document}

\preprint{APS/123-QED}

\title{Fano Resonance and Incoherent Interlayer Excitons \\
in Molecular van der Waals Heterostructures}

\author{Carlos R. Lien-Medrano$^{1}$}
\author{Franco P. Bonafé$^{2}$}%
\author{Chi Yung Yam$^{3}$}
\author{Carlos-Andres Palma$^{4,5}$}
\email[Corresponding author.\\]{palma@iphy.ac.cn}
\author{Cristián G. Sánchez$^{6}$}
\email[Corresponding author.\\]{csanchez@mendoza-conicet.gob.ar}
\author{Thomas Frauenheim$^{3,1,7}$}
\affiliation{%
 $^1$Bremen Center for Computational Materials Science, Universität Bremen, Bremen, Germany
}%
\affiliation{%
 $^2$Max Planck Institute for the Structure and Dynamics of Matter, Hamburg, Germany
}%
\affiliation{%
 $^3$Shenzen JL Computational Science and Applied Research Institute (CSAR), Shenzhen, China
}%
\affiliation{%
 $^4$Institute of Physics, Chinese Academy of Sciences; Beijing, P.R. China.
}%
\affiliation{%
 $^5$Department of Physics \& IRIS Adlershof, Humboldt-Universität zur Berlin, Berlin, Germany
}%
\affiliation{%
 $^6$Instituto Interdisciplinario de Ciencias Básicas, Universidad Nacional de Cuyo, Mendoza, Argentina
}%
\affiliation{%
 $^7$Beijing Computational Science Research Center, Beijing 100193, China
}%

\date{\today}

\begin{abstract}
Complex van der Waals heterostructures from layered molecular stacks are promising optoelectronic materials offering means to efficient, modular charge separation and collection layers. The effect of stacking in the electrodynamics of such hybrid organic–inorganic two-dimensional materials remains largely unexplored, whereby molecular scale engineering could lead to advanced optical phenomena. For instance, tunable Fano engineering could make possible on-demand transparent conducting layers or photoactive elements, and passive cooling. We employ an adapted Gersten-Nitzan model and real time time-dependent density functional tight-binding to study the optoelectronics of self-assembled monolayers on graphene nanoribbons. We find Fano resonances that cause electromagnetic induced opacity and transparency, and reveal an additional incoherent process leading to interlayer exciton formation with a characteristic charge transfer rate. These results showcase hybrid van der Waals heterostructures as paradigmatic 2D optoelectronic stacks, featuring tunable Fano optics and unconventional charge transfer channels.
\end{abstract}

\maketitle

Molecular photosensitizing is a fundamental process in optoelectronics material design. This has led to important advancements in lasing, light emitting diodes and photovoltaic cells \cite{Burschka2013,Vandewal2010,Qian2018,Ramirez2018,Liu2014,Yuan2019}. At the same time, van der Waals (vdW) heterostructures have demonstrated avant-garde optical properties, bearing promise in conception and design of photomodulators, photovoltaic elements \cite{Jariwala2017,Buscema2015,Ermolaev2021}, and bare potential for advanced optical elements, such as antireflection coatings and cooling filters \cite{ElKabbash2021}. In principle, vdW heterojunctions with marginal adsorption and efficiency \cite{Wieghold2016}, may be stacked to achieve photovoltaic efficiencies higher than state-of-the art tandem perovskites \cite{Al-Ashouri1300}.

While gating and doping in two-dimensional (2D) materials is well-known \cite{Wieghold2016,Liu2018,Gobbi2018,Phillipson2016,Samuels2013,Li2013,Coletti2010,Wehling2008,Chen2007,Fleischhauer2005}, the physics of photosensitizing and advanced optical properties have not been fully investigated \cite{Malic2014, Park2021}, especially in the context of molecular vdW heterostructures (MVHs), that is, regular monolayer stacks on 2D materials. Because self-assembled architectures at interfaces grant a high degree of bottom-up control \cite{Barth2005,Ciesielski2010}, a modular and layered approach could be envisaged towards technologically relevant properties, such as photovoltaics, optical switching, lasing and parametric amplification. Just recently for instance, we have shown photoresponsive sensitizers for graphene \cite{Wieghold2016}. The resulting optoelectronic element displayed peak incident photon to current efficiencies of 0.6\%, corresponding to apparent maximum efficiency for the monolayer-thin absorber. Such high responses from monolayers motive further study into the mechanism of photo-induced charge transport and related properties at interfaces.

With accurate computational tools rapidly evolving to handle complex molecular systems, the prototyping of sophisticated sensitized vdW architectonics could become viable. On the other hand, it is well known that the coupling of a discrete localized state to a continuum of states gives rise to a very special phenomenon in optics, the Fano resonance \cite{Limonov2017}. This has been used to describe resonant phenomena in several systems such as photonic crystals, superconductors and plasmonic nanoantennas \cite{Zhou2014,Limonov2000,LukYanchuk2010}, anti-resonance in organic conductors \cite{Gorkov1978} and nonlinear effects on  hybrid molecular-condensed matter systems \cite{Zhang2006} and silicon metasurfaces \cite{Yang2015}. While MVHs have all the necessary elements to show Fano effects in their optical properties, so far this fact has not been investigated.

\begin{figure}[h!]
\includegraphics[]{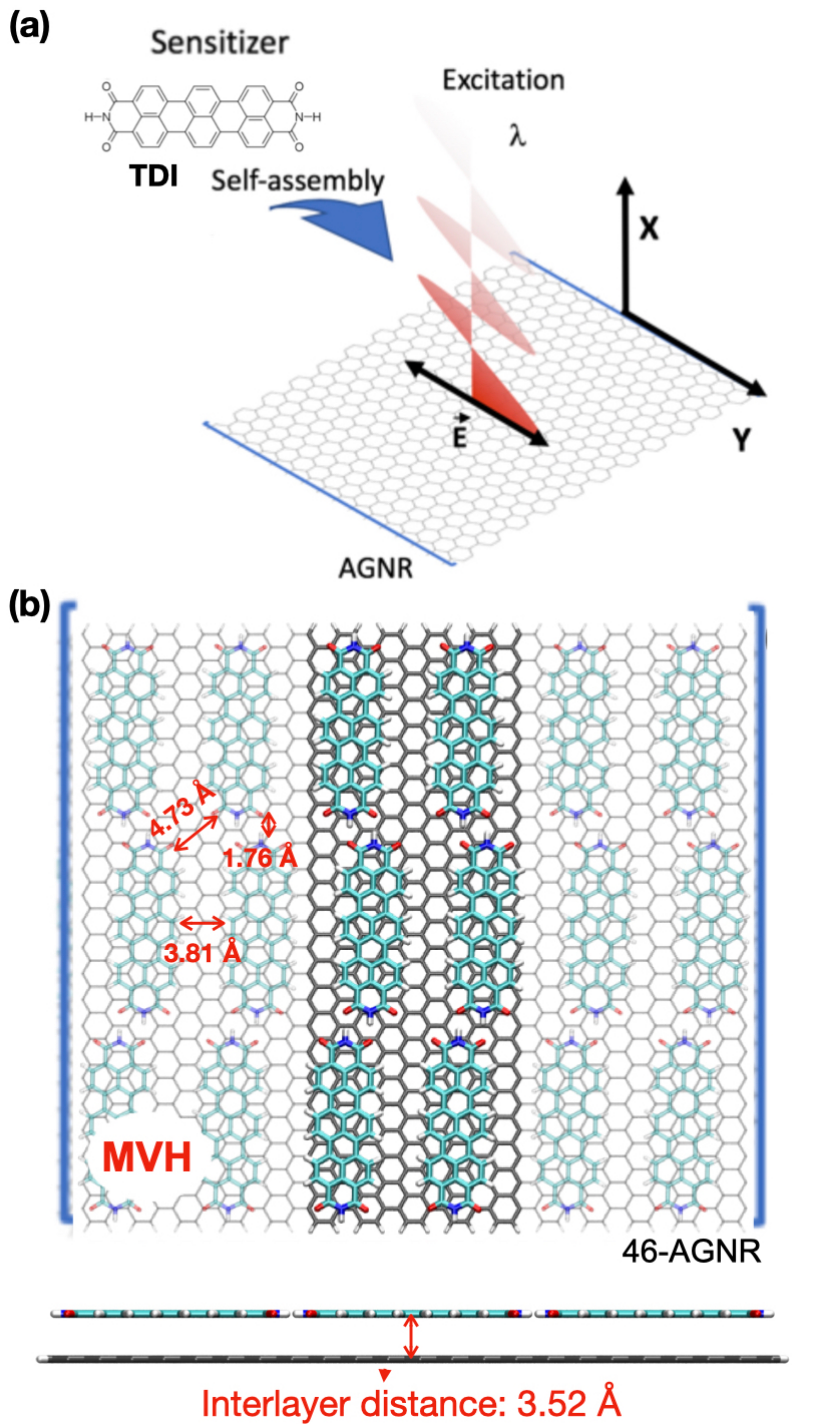}% Here is how to import EPS art
\caption{\label{fig:systems}Schematic diagram of the studied structure. (a) Scheme of the molecular vdW heteroestructures proposed by self-assembly of Terrylene carboxylic diimide (TDI) as monolayer-sensitizer on 1D armchair graphene nanoribbon platform. The external field applied for excitation in the Y axis (nonperiodic direction of the ribbon) is also represented. (b) Monolayer of TDI molecules on top of a 46-atom wide armchair graphene nanoribbon (46-AGNR) as the model heterostructure used for the calculations. The unit cell (shown by thicker bonds) is composed by six TDI molecules interacting by hydrogen bonds on top of the 46-AGNR with a total of 804 atoms. Characteristic distances are depicted in the Figure, particularly, the interlayer distance between monolayer and platform is 3.52 \AA.}
\end{figure}
%FIGURE

To model electronic structure we employ the self-consistent-charge density functional tight binding method (SCC-DFTB). The SCC-DFTB Hamiltonian is based on a second-order expansion of the full Kohn-Sham energy with respect to electronic density fluctuations. This method has proved accurate for the description of quantum properties and electronic structures in large organic systems \cite{Elstner1998, Gaus2014, Christensen2016}.  In particular, we have used the DFTB+ package implementation of the SCC-DFTB method \cite{Hourahine2020} with the pre-computed mio-1-1 set of parameters.\cite{Elstner1998} Here we explore the novel optical features of MVHs in which Fano resonances play a key role. Namely, we have computationally studied the optical properties of broad armchair graphene nanoribbon (AGNR) sensitized by a self-assembled monolayer of rylene dyes (TDI) acceptors by real time time-dependent density functional tight-binding (TD-DFTB) simulations. Our computational framework \cite{Bonafe2020} grants access to the study of nonlinear excitations in systems of thousands of atoms within femto and picosecond timescales. The simulated spectra of the vdW heterostructure shows a remarkable Fano resonance when compared to the isolated components. This is an universal, classical effect of the stacked molecular vdW systems as a result of coupled dipoles, and hence it can be atomically engineered to produce novel, on-demand transparent and particularly IR-transparent photovoltaic elements for solar cells, demonstrating that self-cooling active layers are possible. In addition to the Fano shape for the in-plane response, the out-of-plane response shows non-classical features given by interlayer charge transfer, which can be interpreted as incoherent exciton formation as a decay route of in-plane excitations. Our findings strongly suggest that incoherent interlayer exciton formation is a universal property of self-assembled monolayers at nanosized vdW interfaces which can be tuned at will.

We focus on one representative self-assembly configuration and sensitizer on top of 46-AGNR (Fig. \ref{fig:systems}(a)), using terrylene diimide (TDI) \cite{Wieghold2016}. For the remainder of this work we have considered a self-assembled configuration of TDI as depicted in Fig. \ref{fig:systems}(b). We have simulated the absorption spectra via reduced single-electron density matrix propagation within TD-DFTB, as explained in ref. \citenum{Bonafe2020} (for more details on the computational methods please refer to the Materials and methods section of the supporting information).
We find that the broad armchair graphene nanoribbon 46-AGNR is a small-gap (0.1 eV) semiconductor with absorptivity near 5\% 1.0-3.5 eV along the aperiodic [$\Gamma$-M] direction (see supporting information Fig. S1). 
These findings are in good agreement with previous TD-DFT\cite{Lian2018} absorption spectra calculations for AGNR with different widths.
The absorption spectrum between 1 and 3.5 eV is also in good agreement with the results reported by Yang et al.\cite{Yang2009} for graphene with and without electron-hole interactions (since the electron-hole interactions do not affect the absorption in this region but at higher energies).
These results guarantee that the 46-AGNR can be used as a graphene-like platform for this study, mimicking all of the relevant properties of the fully periodic 2D system, which could not be treated within our method.

\begin{figure*}
\includegraphics[]{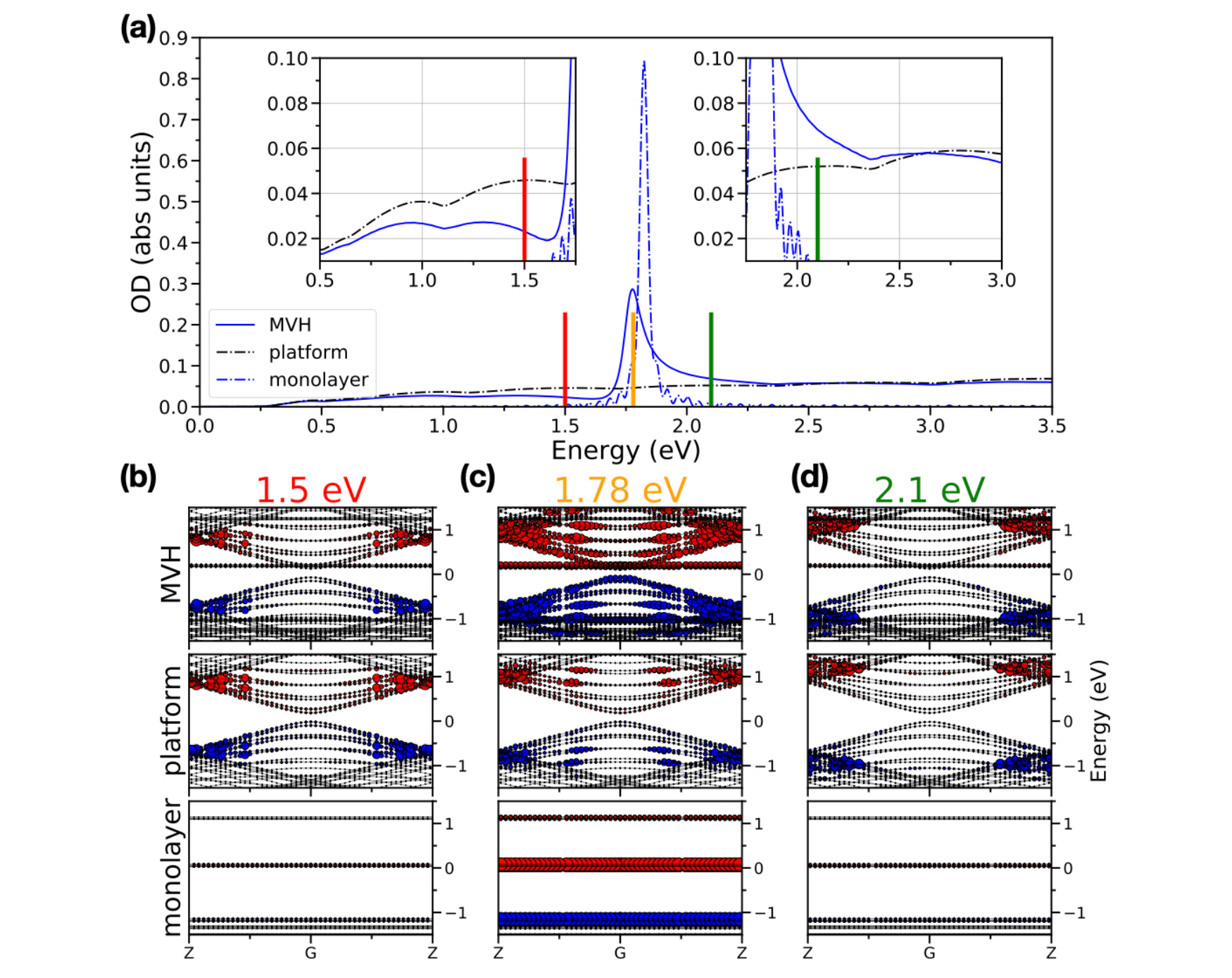}% Here is how to import EPS art
\caption{\label{fig:spec}Theoretical absorption spectrum of MVH and band population analysis of the Fano features. (a) Simulated absorption spectra for the MVH studied and its separate components (ribbon platform and dyes monolayer). Note that the fluctuations around the absorption peak of the monolayer 
bear no physical meaning, as they are due to the mathematical form of the Fourier transform of a Dirac delta, which is evident by the absence of an intrinsic lifetime of the molecular transition within this level 
of theory. (b), (c) And (d) present the band populations under continuous excitation at three different energies, 1.5, 1.78 and 2.1 eV respectively. Red (blue) circles denote the increase (decrease) of electron occupation and the circle size is proportional to the change of population. In order to see the differences, all cases show first the composed heterostructure (MVH) and then the separate components (platform and monolayer). The excitations were performed using a continuous laser perturbation tuned with the energy of interest. For all the band populations the snapshot frame was taken at 96 fs of the electron dynamics simulation.}
\end{figure*}

The calculated spectra reveal absorption suppression or enhancement upon adsorption of the sensitizers around the dye energy excitation, when compared to the sum of the absorption of the separate components (Fig. \ref{fig:spec}(a)). An absorption maximum of 692 nm is found for the monolayer on the AGNR platform (cf. 735 nm dye excitation on graphene \cite{Wieghold2016}). Removal of the underlying graphene platform blueshifts the absorption peak position by 13 nm, to 679 nm.

The spectral features of the combined system can be resolved in k-space, as depicted in Figs. \ref{fig:spec}(b), (c) and (d). Here, we show a momentum resolved snapshot of the time-dependent electron population distribution at 96 fs, projected on the ground state bands for 3 different excitation energies: 1.5 eV (\ref{fig:spec}(b), suppression/transparency), 1.78 eV (\ref{fig:spec}(c), dye excitation), and 2.1 eV (\ref{fig:spec}(d), enhancement/opacity). The MVH as well as its separate components are considered for the plots. Red (blue) circles denote the increase (decrease) of electron occupation, and the circle size is proportional to the change of population. The fundamentals of this method can be found in refs \cite{Bonafe2020,Lian2018}. The absorption suppression of the heterostructure at 1.5 eV agrees with a reduction of population change with respect to the free platform (Fig \ref{fig:spec}(b)). The opposite behaviour is depicted in Figure \ref{fig:spec}(d), where the heterostructure shows an increase of the population change with respect to the free platform, in concordance with the absorption enhancement at 2.1 eV. Furthermore, the monolayer electronic population change at 1.78 eV shown in Fig. \ref{fig:spec}(c) is clearly quenched in the heterostructure system with respect to the free standing monolayer. 
%the monolayer electronic population change is quenched in the composite system with respect to the freestanding case
The population change involving different states than those of the separate system in Fig. \ref{fig:spec}(c) is also evidence of novel interlayer transitions (vide infra). It is also proven here that the suppression and enhancement effect on the absorption depends on the coverage degree of the platform as shown in Fig. S2 for a range of TDI submonolayers.

\begin{figure}
\centering
\includegraphics[]{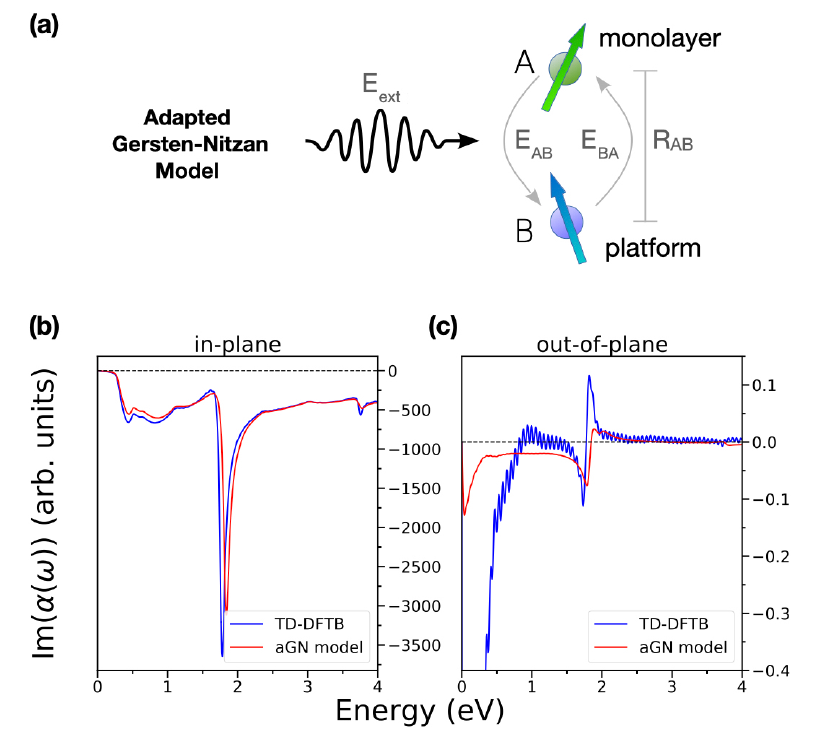}% Here is how to import EPS art
\caption{\label{fig:polarizabilities}Adapted Gersten-Nitzan Model and Comparison with TD-DFTB results. (a) Simple model representation of two point dipoles interacting with each other via dipolar field and with an external field. (b) In-plane imaginary component of the Polarizability tensor obtained from the adapted GN model and TD-DFTB simulations for the MVH. (c) Out-of-plane imaginary component of the Polarizability tensor obtained from the adapted GN model and TD-DFTB simulations of MVH. The imaginary part of the in-plane component is related to the absorption spectrum presented in Fig. \ref{fig:spec} and clearly shows the Fano shape.}
\end{figure}

The monolayer electronic structure can be described by molecular energy levels (without dispersion), while the platform has a wide and relatively flat (or quasi-linear) absorption over a broad range of energies from the UV to the IR region which include those of the monolayer excitation in the visible. These two facts are textbook conditions for a Fano resonance \cite{Stefanucci2013}. To shed light on the origin of the Fano resonance present in Fig. \ref{fig:spec}(a), we used an adapted Gersten-Nitzan (aGN) model to calculate the polarizability of the heterostructure from the polarizabilities of the separate components. The well-known Gersten-Nitzan model \cite{Gersten1980} has been used successfully over 40 years in the realm of surface-enhanced Raman spectroscopy to calculate radiative and non-radiative decay rates of molecules adsorbed on metals \cite{Barzan2016,Cheng2011,Yin2011}. The basic setup considered for the model are two point dipoles, each of them coupled to the external field, and to the dipolar field generated by the other system up to first order, neglecting image field effects, as depicted in Fig. \ref{fig:polarizabilities}(a).

The effect of the dipolar coupling between the systems in such setup can be described by an effective polarizability (see supporting information for formal definition), which for system A reads:
\begin{equation}
    \alpha^{\mathrm{eff}, A} = \alpha^A(\omega)+ \alpha^A(\omega) D \alpha^B(\omega) ,
   \label{eq:eff_pol_a}
\end{equation}
where
\begin{equation}
    D = \frac{1}{4\pi\epsilon_0 R_{AB}^3}\left(
\begin{array}{cc}
2 & 0 \\
0 & 1
\end{array}\right)
\end{equation}
is the geometric factor for two point dipoles separated at (interlayer) distance $R_{AB}$ (see Fig. \ref{fig:polarizabilities}(a)), as thoroughly explained in the supporting information section 1.2. Exchanging indices A and B gives the equivalent formula for system B. The total polarizability is the sum of both.

The polarizabilities used as input for the model were calculated for the separate systems in vacuum using real time TD-DFTB (see Fig. S3). The imaginary part of the in-plane ({\bf ip}) (y-axis response from y-polarized field) polarizability calculated from the aGN model (Fig. \ref{fig:polarizabilities}(b)) shows a Fano resonance around 1.78 eV. When a broadening of 13 fs is added to the monolayer polarizability (\textit{vide infra}), the curve obtained agrees quantitatively with the one calculated for the full combined system using TD-DFTB (same Fig. \ref{fig:polarizabilities}(b)). This proves that the {\bf ip} transparency/enhancement arises fundamentally from a dipolar coupling between the systems. However, the out-of-plane ({\bf oop}) (y-axis response for x-polarized field) polarizability calculated from the aGN model (Fig. \ref{fig:polarizabilities}(c)) agrees only qualitatively with the TD-DFTB one. 

\begin{figure}
\includegraphics[]{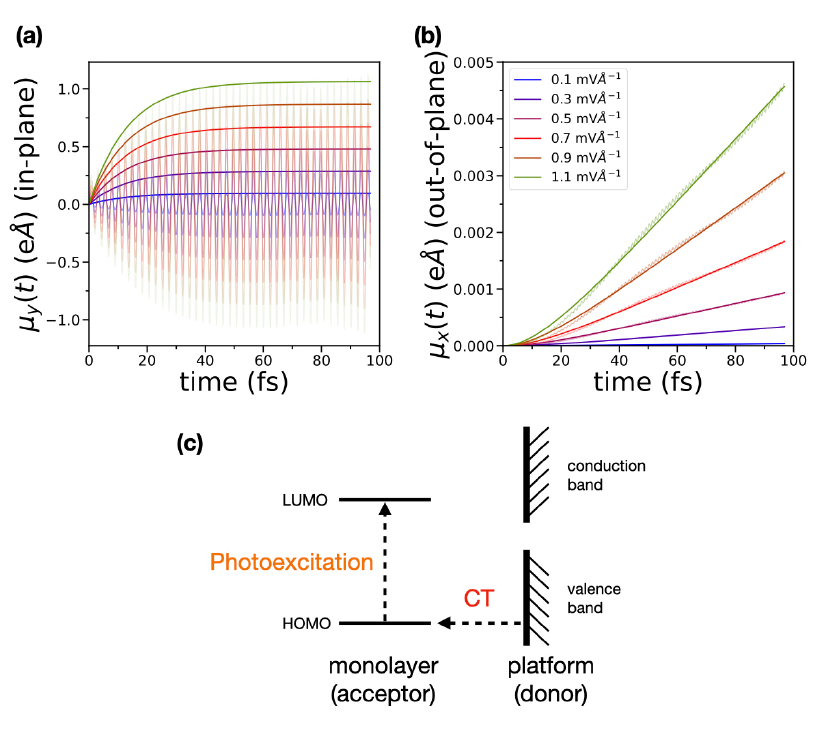}% Here is how to import EPS art
\caption{\label{fig:mu}Time dependent analysis of dipole moments with respect to the field strength $E_0$. (a) In-plane dipole moment ($\mu_y$) vs. time for different $E_0$ values. (b) Out-of-plane dipole moment ($\mu_x$) vs. time for different $E_0$ values. (c) Schematic representation of the two processes involved, photoexcitation of dye molecules and charge transfer from the ribbon platform to the molecules (exciton formation). The mathematical expressions used for fitting dipole curves in A and B are presented in the text. All excitations were performed at 1.78 eV (dye excitation energy).}
\end{figure}

In order to gain insight into the {\bf oop} process, we studied the electron dynamics upon photoexcitation of the dye resonance in the {\bf ip} direction. Figure \ref{fig:mu}(a) shows the {\bf ip} dipole moment $\mu_y$(t) for different field strengths. The {\bf ip} dipole moment grows linearly during the first steps (up to approximately 15 fs) and then reaches a stationary state showing a saturation value that depends on the field strength. The amplitudes of the different $\mu_y$(t) were fitted following the expression:
\begin{equation}
 \mu_y^{\mathrm{fit}} = \mu_y^{\mathrm{sat}}(1-e^{-\frac{t}{\tau}}),
    \label{eq:mu_y}
\end{equation}
where $\mu_y^{\mathrm{sat}}$ is the saturation dipole moment, which is directly proportional to the field strength $E_0$,  and $\tau$ is the characteristic lifetime of the saturation process with a constant value of 13 fs. In the linear response regime, a linear growth of the dipole moment in time would be expected. The observed saturation evidences a secondary process after the photoexcitation of the dye molecules. Fig. \ref{fig:mu}(b) shows the {\bf oop} dipole moment $\mu_x$ as a function of time for the same dynamics in Fig. \ref{fig:mu}(a).
The $\tau$ value obtained from the fitted curves was used as the broadening of the {\bf ip} molecular polarizability in \ref{fig:polarizabilities}(b). This broadening is essential for the quantitative agreement between the aGN model and the full quantum results and shows the existence of a decay channel for the molecular excitation. 

The appearance of an {\bf oop} dipole moment indicates the cause of the decay of the excitation is interlayer charge transfer (CT) from the platform to the monolayer. The dipole moment $\mu_x$(t) was fitted to the following expression (the time integral of equation \ref{eq:mu_y}):
\begin{eqnarray}
    \displaystyle \mu_x^{\mathrm{fit}} = \gamma \mu_y^{\mathrm{sat}}(t-\tau (1-e^{-\frac{t}{\tau}})),
    \label{eq:mu_x}
\end{eqnarray}
where $\mu_y^{\mathrm{sat}}$ and $\tau$ are the same as in equation \ref{eq:mu_y} and $\gamma$ represents the charge leaking between both systems and depends linearly on $E_0$ (so does $\mu_y^{\mathrm{sat}}$). Both processes, excitation (equation \ref{eq:mu_y}) and CT (equation \ref{eq:mu_x}) are
schematically represented in Fig. \ref{fig:mu}(c). The value of charge transferred $q_{CT}^{fit}$ can be obtained from eq. \ref{eq:mu_x} dividing by the interlayer distance between the monolayer and the platform:
\begin{eqnarray}
    \displaystyle
    q_{CT}^{fit}=\mu_x^{fit} / R_{AB}.
    \label{eq:q_ct}
\end{eqnarray}

Figure S4 shows that the $\Delta q(t)$ obtained from the simulation agrees with the results from  equation \ref{eq:q_ct} with $R_{AB}$ approximately equal to the interlayer distance shown in Fig. \ref{fig:systems}(b). The CT process explains the differences from Fig. \ref{fig:polarizabilities}(c),  as the aGN model cannot account for charge leaking.  Hence, continuous wave excitation of the {\bf ip} mode decays into incoherent interlayer excitons in which a net {\bf oop} dipole moment is caused by tunnelling of electrons from the platform to the monolayer, which are revealed by a linearly increasing, non-oscillating dipole moment in the {\bf oop} direction. Similar incoherent excitons have been previously described in transition metal dichalcogenides based van der Waals heterostructures through tunneling between energetically resonant states resulting in a transfer of holes \cite{Ovesen2019}.
Upon closer inspection of the band diagrams in Fig. \ref{fig:spec}(c) (see Fig. S5), avoided crossings can be seen for the HOMO and HOMO-1 orbitals of the monolayer. The avoided crossings explain the direction of charge transfer, as only the occupied frontier orbitals from the molecule are mixed with the substrate, establishing a clear channel for charge leakage from the platform into partially depopulated monolayer states.
Once the linear regime has been established in the {\bf oop} dipole moment charge transfer current reaches a steady state. This current is linear with illumination intensity (see figure S6) with a slope (photoresponse) of 1.3 mA W$^{-1}$.

In summary, we have found evidence of induced transparency and opacity of the substrate arising from a Fano spectral shape upon formation of the MVH architecture. This spectral change  is a general feature of these types of systems, and can already be captured by a simple module of electrostatically interacting dipoles, which is valid even for the interplanar distances at equilibrium. This simple model cannot capture, however, the bandwidth of the response calculated from a fully atomistic, quantum-dynamical treatment of the system which is given by the lifetime of a new decay channel. Our time-resolved carrier dynamics simulations suggest that such a process could be understood as the formation of an incoherent interlayer exciton.
These findings are relevant for the field of molecular layers at well-defined interfaces, atomic-scale vdW metamaterials and precision molecular devices, to mention a few. For instance, the Fano profile contains information about the polarizability, molecular geometry at interfaces and distance between molecules and substrate, so that optical studies could complement synchrotron X-ray absorption and diffraction methods for interfacial structure elucidation. Regarding vdW molecular metamaterials, the adsorption spectrum of multiple stacked molecular layers and 2D materials could be estimated with the AGN approach to engineering the spectral shape of optical transparency or enhancement.  Finally, our findings open a path for improved design of modular multilayer organic PV devices, by e.g. choosing molecular materials which absorb in-plane and feature out-of-plane dipoles for enhanced interlayer charge transfer.

C.R.L.-M. acknowledges the support from the DFG-RTG2247 grant (RTG-QM3 Program) for the post-doctoral fellowship and the high-computing resources of the BCCMS, University of Bremen. F.P.B. acknowledges financial support from the European Union’s Horizon 2020 research and innovation programme under the Marie Skłodowska-Curie grant agreement no. 895747 (NanoLightQD). C.-A.P. acknowledges support from the NSFC China (nos. 51325204 and 51761135130), CAS Bureau of Frontier Sciences and Education no. QYZDB-SSW-SLH038 and CAS Strategic Priority Research Program XDB33000000 the Cluster of Excellence ``Matters of Activity'' funded by the DFG - EXC 2025 - 390648296 and Alexander von Humboldt Foundation. C.G.S. acknowledges financial support by Consejo Nacional de Investigaciones Científicas y Técnicas (CONICET) through Grant PIP 112-2017-0100892CO and Agencia Nacional de Promoción Científica y Tecnológica (ANPCYT) through grant PICT-2017-1605. The authors would like to aknowledge the previous experimental work from Sarah Wieghold and Juan Li that motivates the current work and the insightful discussions with Dr. Yair Litman.

\bibliography{ms}

\end{document}

\preprint{APS/123-QED}

\title{Fano Resonance and Incoherent interlayer excitons \\
in molecular van der Waals heterostructures}

\author{Carlos R. Lien-Medrano$^{1}$}
\author{Franco P. Bonafé$^{2}$}%
\author{Chi Yung Yam$^{3}$}
\author{Carlos-Andres Palma$^{4,5}$}
\email[Corresponding author.\\]{palma@iphy.ac.cn}
\author{Cristián G. Sánchez$^{6}$}
\email[Corresponding author.\\]{csanchez@mendoza-conicet.gob.ar}
\author{Thomas Frauenheim$^{3,1}$}
\affiliation{%
 $^1$Bremen Center for Computational Materials Science, Universität Bremen, Bremen, Germany
}%
\affiliation{%
 $^2$Max Planck Institute for the Structure and Dynamics of Matter, Hamburg, Germany
}%
\affiliation{%
 $^3$Shenzen JL Computational Science and Applied Research Institute (CSAR), Shenzhen, China
}%
\affiliation{%
 $^4$Institute of Physics, Chinese Academy of Sciences; Beijing, P.R. China.
}%
\affiliation{%
 $^5$Department of Physics \& IRIS Adlershof, Humboldt-Universität zur Berlin, Berlin, Germany
}%
\affiliation{%
 $^6$Instituto Interdisciplinario de Ciencias Básicas, Universidad Nacional de Cuyo, Mendoza, Argentina
}%
\affiliation{%
 $^7$Beijing Computational Science Research Center, Beijing 100193, China
}%

% \date{\today}

\maketitle

\section{Computational methods}

\subsection{Real time time-dependent density tigth-binding}
We have used the DFTB+ package \cite{Hourahine2020}, an implementation of the DFTB method, to obtain the GS Hamiltonian (H\textsubscript{GS}) and overlap matrix (S). Using the H\textsubscript{GS} and the S matrix, we computed the the initial GS reduced single-electron density matrix ($\rho$). The mio-1-1 DFTB parameters set was employed to obtain the electronic structure of all the structures presented in this work. 
In order to describe the electronic dynamics of the systems under study we need to extend the DFTB method to the time-domain (TD-DFTB). On the basis of a real-time propagation of $\rho$ under the influence of a time-varying external field, we can obtain excited-state properties of the systems. This propagation is achieved through the numerical integration of the 
Liouville-von Neumann equation of motion in the non-orthogonal basis:
\begin{equation*}
 \dot{\rho} = -\mathrm{i} (S^{-1} H \rho - \rho H S^{-1}).
\label{eq:eomrho}
\end{equation*}
For more details on the theoretical method and its
computational implementation, please refer to Bonafé et al. \cite{Bonafe2020}.  By applying a perturbation in the shape of a Dirac delta to the Hamiltonian, within the linear response regime, we can obtain the absorption spectra of the sytems as follows. The dipole moment is given by:
\begin{equation*}
\boldsymbol{\mu}(t)=\int_{0}^{\infty} \boldsymbol{\alpha}(t-\tau)E(\tau)d\tau,
\end{equation*}
where $\boldsymbol{\alpha}(t-\tau)$ is the polarizability along the axis over which the external field $\mathbf{E}(\tau)$ is applied. After the deconvolution of the applied electric field, the frequency dependent polarizability $\alpha$ can be obtained:
\begin{equation*}
\label{alpha}
\boldsymbol{\alpha}(\omega)=\frac{\boldsymbol{\mu}(\omega)}{E(\omega)}
\end{equation*}
The imaginary part of the polarizability is proportional to the Absorption spectrum.

For the study of the charge transfer processes, electron dynamics were triggered by the application of a continuous laser-type (sinusoidal) perturbation in tune with the excitation energy of interest. 

% \subsubsection*{Note}
% DFTB methodology employed (and by construction, the TD-DFTB implementation) ignores the many-body effects of excited states and exciton states. State-of-the-art techniques currently employ a posteriori corrections to include these two effects, often in the frameworks of GW quasiparticle approximation and BSA exciton binding approximation. We have previously suggested that for nanographenes, the corrections are approximate canceled as DFTB phenomenologically predicts the optical gap of nanographenes. The accurate predictions of bathochromic shifts, quenching and photovoltaic efficiencies of TDI on graphene, strongly suggests that the TD-DFTB methology can be employed to accurately engineer vdW interstack excitons. These predictions are expected to deviate for highly correlated materials but are highly plausible for systems where many-body effects are negligible. 

\newpage
\subsection{Derivation of the adapted Gersten-Nitzan model}

Let $A$ and $B$ be two coupled point-dipoles, as depicted in Figure 3 (a) of the main manuscript. They are separated in the $x$ direction by a  distance $R_{AB}$. Both systems are coupled to an external time-dependent electric field $\ext(t)$, which can have any polarization direction within the $(x,y)$ plane. Moreover, both systems are affected by the dipolar electric field generated by the other system. In other words, being $\ef_A$ and $\ef_B$ the effective electric fields acting on both systems,

\begin{align*}
\ef_A (t) = \ext(t) + \ef_{BA}(\rvec_A,t) \\
\ef_B (t) = \ext(t) + \ef_{AB}(\rvec_B,t) \\
\end{align*}

where $\ef_{BA}$ and $\ef_{AB}$ are the electric field generated by $B$ acting on $A$, and the one generated by $A$ acting on $B$, respectively, defined as:

\begin{align}
\ef_{BA}(R_A,t) = \intmenosinf{t} \chitilde_B(\rvec_A, t, t') \ext(t') dt' 
\label{eq:ebachitilde1} \\
\ef_{AB}(R_B,t) = \intmenosinf{t} \chitilde_A(\rvec_B, t, t') \ext(t') dt'
\label{eq:ebachitilde2}
\end{align}

Here, $\chitilde$ is a {\bf dipole-field response function} that is such that when multiplied by the external field and integrated yields the local dipole electric field generated by the system. It has a connection with the dipole susceptibility, which will be explored below. We start with the formula for the electric field of a perfect dipole, and being $\vecmu_I$ be the dipole moment of system $I$:

\begin{align}
\ef_{BA} = \elfactor \frac{3(\vecmu_B \cdot \hat{R}_{BA})\hat{R}_{BA} - \vecmu_B}{R_{BA}^3}
\label{eq:ebadipole}
\end{align}

As usual, $\rvec_{BA} = \rvec_A - \rvec_B$. By construction of the model, $\rvec_{BA} = - R_{AB} \hat{x}$, which means $\hat{R}_{BA} = - \hat{x}$; and $\rvec_{AB} = R_{AB} \hat{x}$, ergo $\hat{R}_{AB} = \hat{x}$. By inserting the position vectors into eq. \ref{eq:ebadipole}, we get:

\begin{align}
\ef_{BA} = \elfactor \frac{2 \mu_{B_x} \hat{x} - \mu_{B_y} \hat{y} }{R_{AB}^3} \label{eq:efbamub}
\end{align}

Now, by definition of the dipole response function $\chi$, and neglecting image field effects,

\begin{align}
    \vecmu_B(t) = \vecmu_B^0 + \intmenosinf{t} \chi^B(t,t') \ext(t') dt'
    \label{eq:mub}
\end{align}

Inserting \ref{eq:mub} into \ref{eq:efbamub}:

\begin{align}
\ef_{BA} &=\intmenosinf{t} \elfactorrcubo
\begin{pmatrix}
2 & 0 \\
0 & -1
\end{pmatrix} \cdot \chi^B(t,t') \cdot \ext(t')  dt' \label{eq:ebachib}
\end{align}

which by comparison with \ref{eq:ebachitilde1}, allows us to define $\chitilde$ for system $B$:

\begin{align}
\chitilde_B &=\elfactorrcubo \begin{pmatrix}
2 & 0 \\
0 & -1
\end{pmatrix} \chi^B(t,t') \label{eq:chitildeb}
\end{align}

As everything is symmetric under the exchange of labels $A\rightarrow B$ and $B \rightarrow A$, we get now for $\ef_{AB}$ and $\chitilde_A$ the following:

\begin{align}
    \ef_{AB} &=\intmenosinf{t} \elfactorrcubo \begin{pmatrix}
2 & 0 \\
0 & -1
\end{pmatrix} \chi^A(t,t') \cdot \ext(t') ) dt' \label{eq:ebachia} \\
\chitilde_A &=\elfactorrcubo \begin{pmatrix}
2 & 0 \\
0 & -1
\end{pmatrix}  \chi^A(t,t') \label{eq:chitildea}
\end{align}

Our goal is to get the polarizability of the combined system under the effect of the external field and of the dipolar field induced in the neighbour system, using as inputs the polarizabilities of the isolated systems. Then, we need to calculate both terms of the total dipole $\vecmu (t) = \vecmu_A(t) + \vecmu_B(t)$. Starting with $\vecmu_A(t)$, we have the the external field and the induced dipolar field terms:

\begin{align}
\vecmu_A(t) = \intmenosinf{t} dt' \chi^A(t,t') \ext(t') +  \intmenosinf{t} \intmenosinf{t'} dt' dt'' \chi^A(t,t') D  \chi^B(t',t'') \ext(t'') \label{eq:muachiachib}
\end{align}

where  $D := \elfactorrcubo \begin{pmatrix}
2 & 0 \\
0 & -1
\end{pmatrix} $ is the geometrical factor. Calculating the Fourier transform of \ref{eq:muachiachib}:

\begin{align}
\vecmu_A(\omega) = [\alpha^A(\omega)+ \alpha^A(\omega) D \alpha^B(\omega)] \ef(\omega) \label{eq:muafinal}
\end{align}

And equivalently for $B$:

\begin{align}
\vecmu_B(\omega) = [\alpha^B(\omega) + \alpha^B(\omega) D \alpha^A(\omega)] \ef(\omega) \label{eq:mubfinal}
\end{align}

These expressions suggest a natural definition of an effective polarizability, which can be written as:

\begin{align}
   \alpha^{\mathrm{eff}, A} = \alpha^A(\omega)+ \alpha^A(\omega) D \alpha^B(\omega) 
   \label{eq:eff_pol_a}\\
   \alpha^{\mathrm{eff}, B} = \alpha^B(\omega)+ \alpha^B(\omega) D \alpha^A(\omega)
 \label{eq:eff_pol_b}
\end{align}

Which is the formula used to calculate the polarizability components shown in Figure 3 of the main manuscript.

\newpage

\section{Figs. S1 to S6}
\vspace{1cm}

\begin{figure}[h!]
    \centering
    \includegraphics[width=0.7\linewidth]{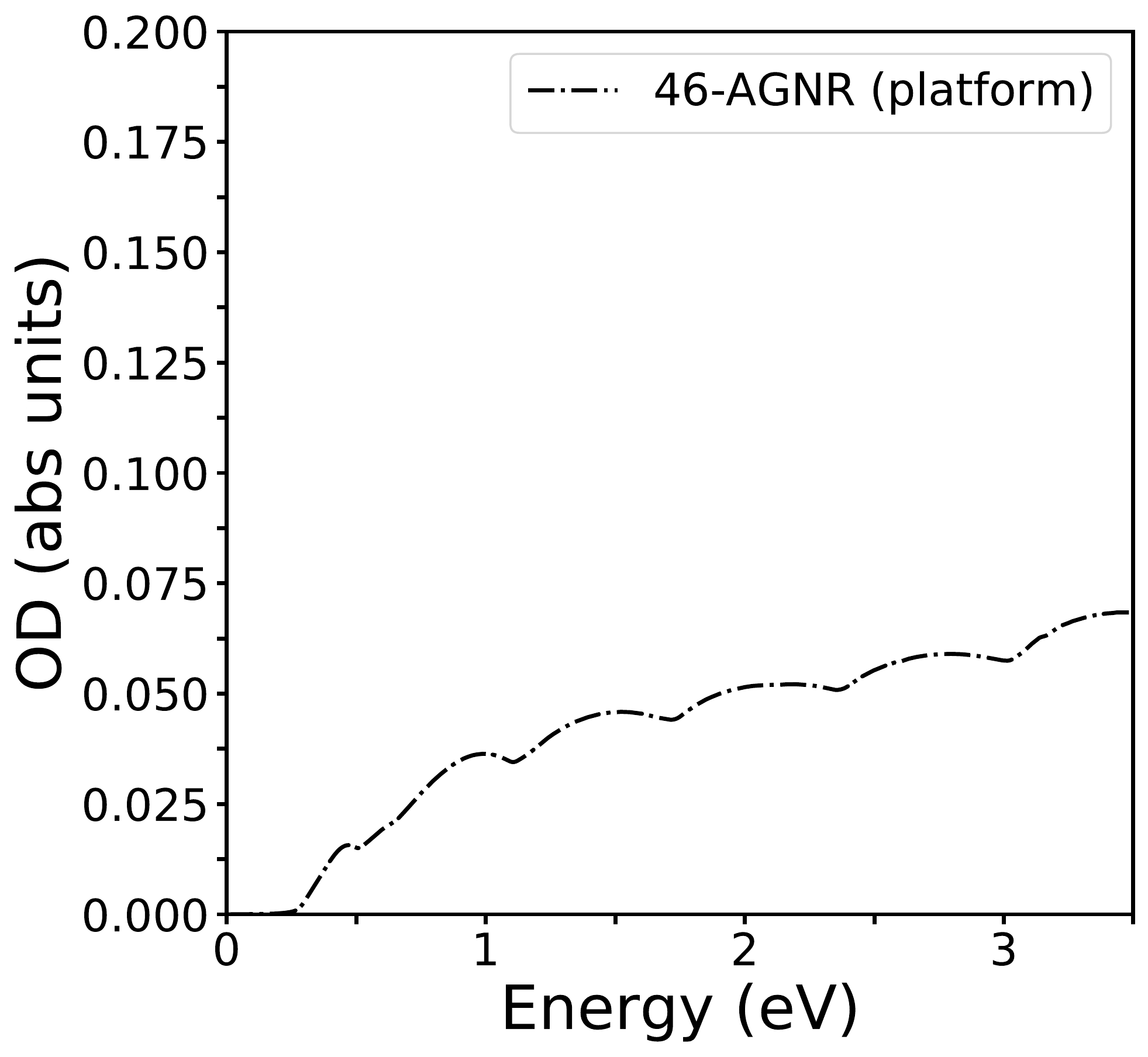}
    % \caption{Caption}
    % \label{fig:S_1}
\end{figure}

\noindent
{\bf{Fig. S1. Absorption spectrum of 46-AGNR.}}\\
The figure shows the absorption spectrum of the 46 atom width Armchair graphene nanoribbon (46-AGNR) used as the platform in the van der Waals molecular heterostructure presented in this work.

\newpage

% \begin{figure}[hb!]
%     \centering
%     \includegraphics[width=0.7\linewidth]{Fig_S2.pdf}
%     % \caption{Caption}
%     % \label{fig:S_2}
% \end{figure}
% \noindent
% {\bf{Fig. S2 Absorption spectrum of broadband absorber OMe-CGNR.}}\\
% The figure shows the absorption spectrum for molecule 2 of the main manuscript. 

% \newpage
% % \begin{figure}[hb!]
% %     \centering
% %     \includegraphics[width=0.7\linewidth]{Fig_S3.pdf}
% %     \caption{Caption}
% %     \label{fig:S_3}
% % \end{figure}
% \vspace{2cm}
% {\bf{Fig. S3 In progress.}}

\newpage

\begin{figure}[hb!]
    \centering
    \includegraphics[width=1.\linewidth]{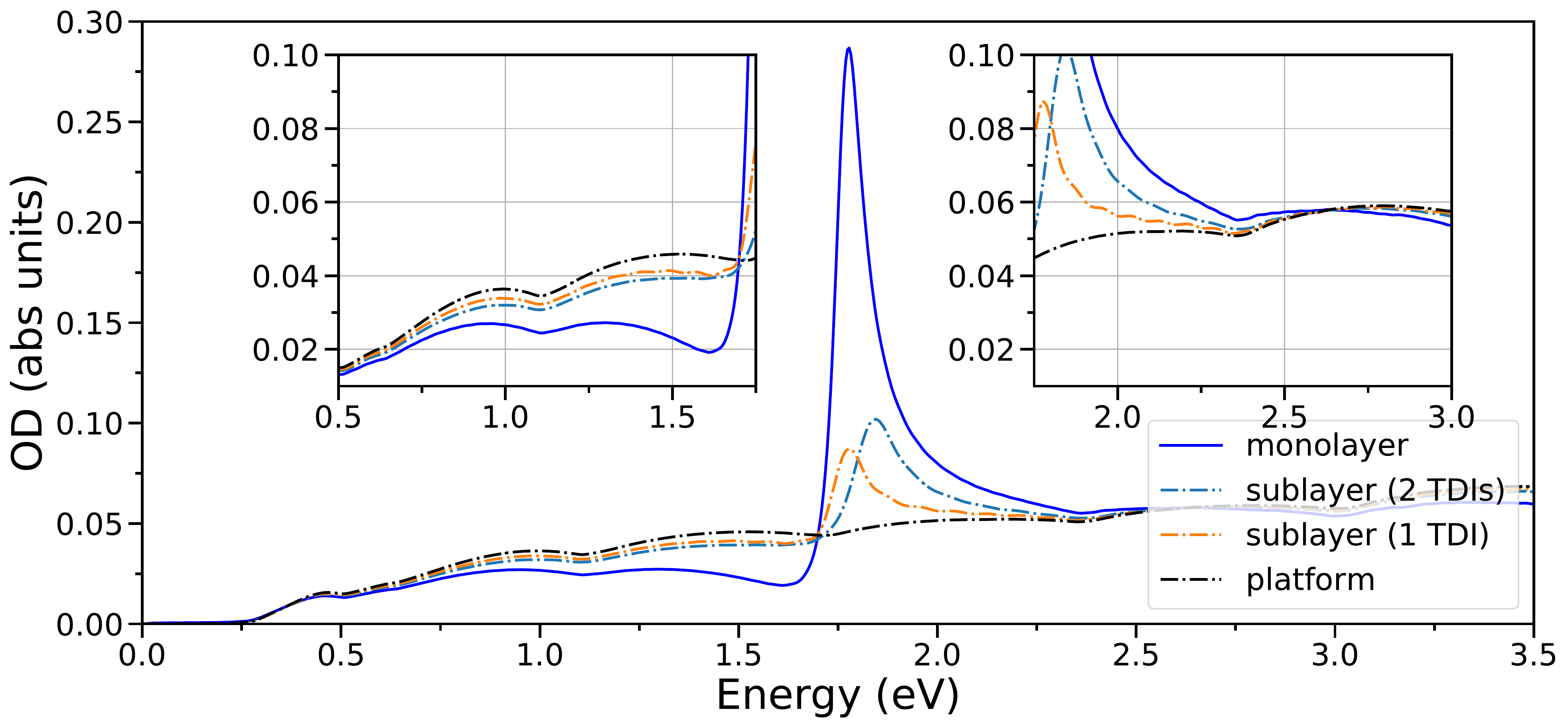}
    % \caption{Caption}
    % \label{fig:S_4}
\end{figure}
{\bf{Fig. S2 Coverage absorption dependency.}}
Absorption spectrum of different sublayers of TDI molecules on top of the ribbon platform. As a reference, in blue is plotted the same monolayer spectrum as in the manuscript (6 TDI molecules in the unit cell).

\newpage

\begin{figure}[hb!]
    \centering
    \includegraphics[width=1.\linewidth]{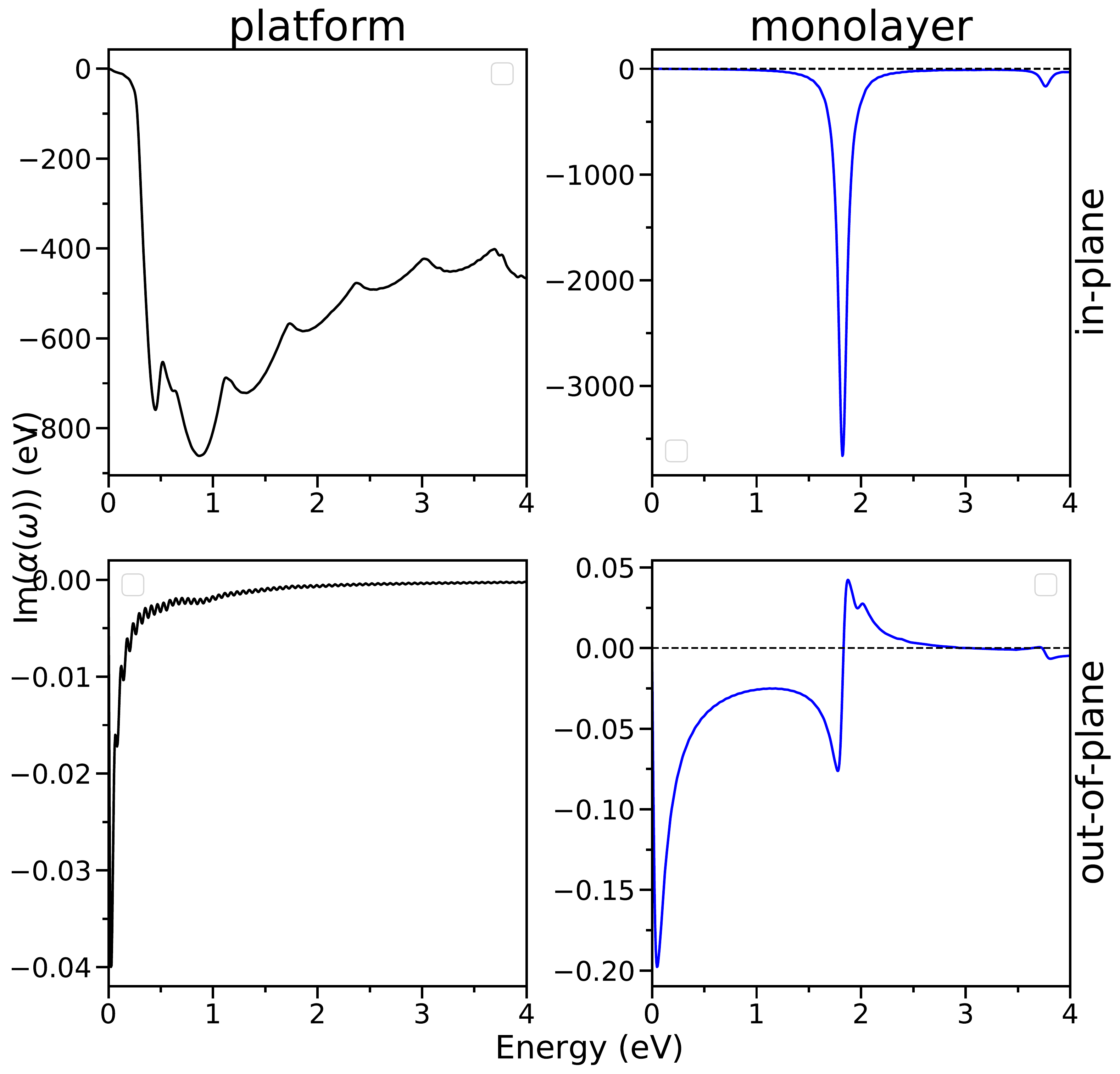}
    % \caption{Caption}
    % \label{fig:S_5}
\end{figure}
{\bf Fig. S3. Individual polarizabilities for system 1a.}
({\bf A}) In-plane and out-of-plane polarizabilities for the platform. ({\bf B}) In-plane and out-of-plane polarizabilities for the monolayer of TDI.

\newpage

\begin{figure}[hb!]
    \centering
    \includegraphics[width=.7\linewidth]{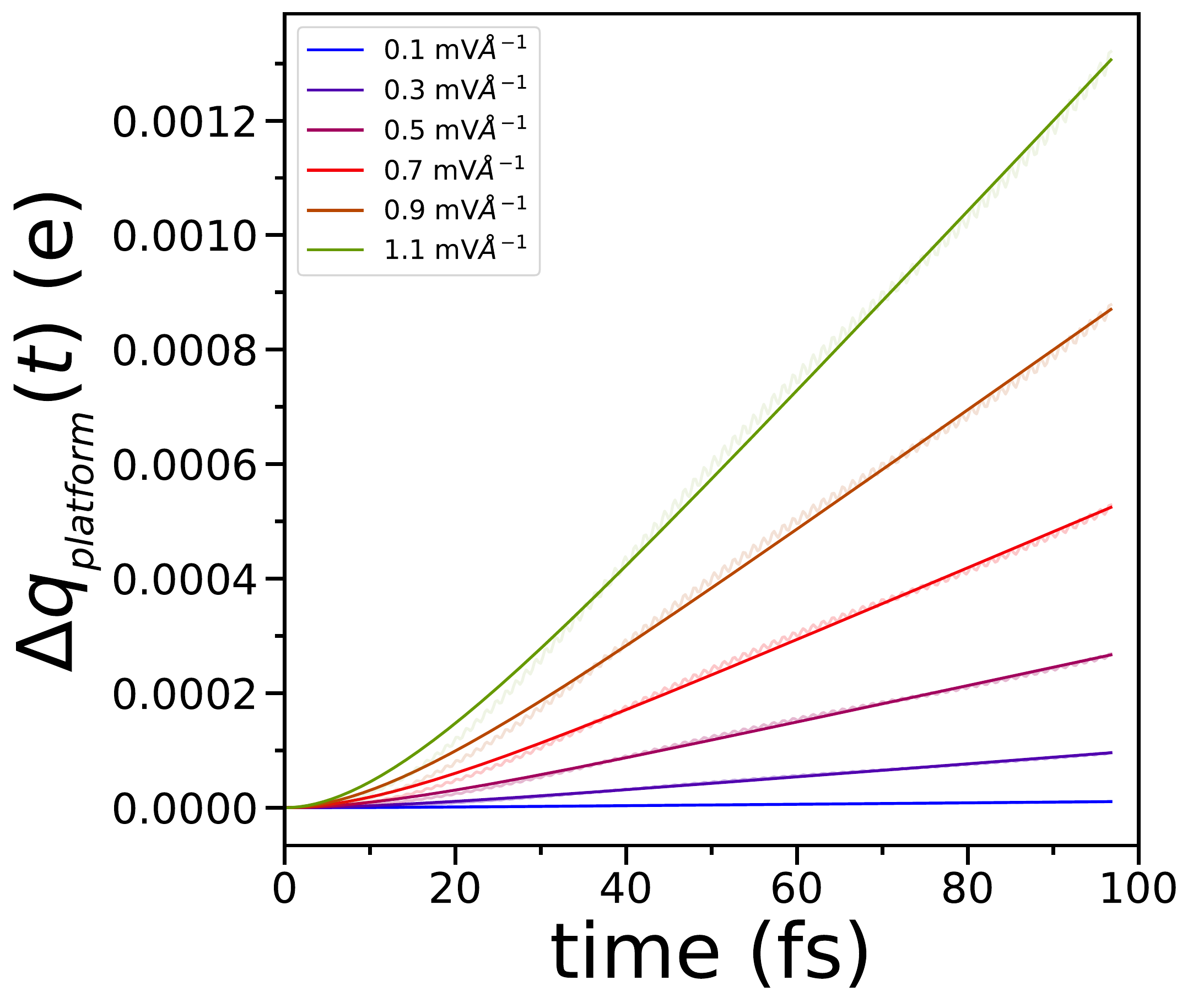}
    % \caption{Caption}
    % \label{fig:S_5}
\end{figure}
{\bf Fig. S4. Charge transfer}
$\Delta q(t)$ of the platform for different field strengths as obtained from the dynamical simulations. 

\newpage

\begin{figure}[hb!]
    \centering
    \includegraphics[width=.6\linewidth]{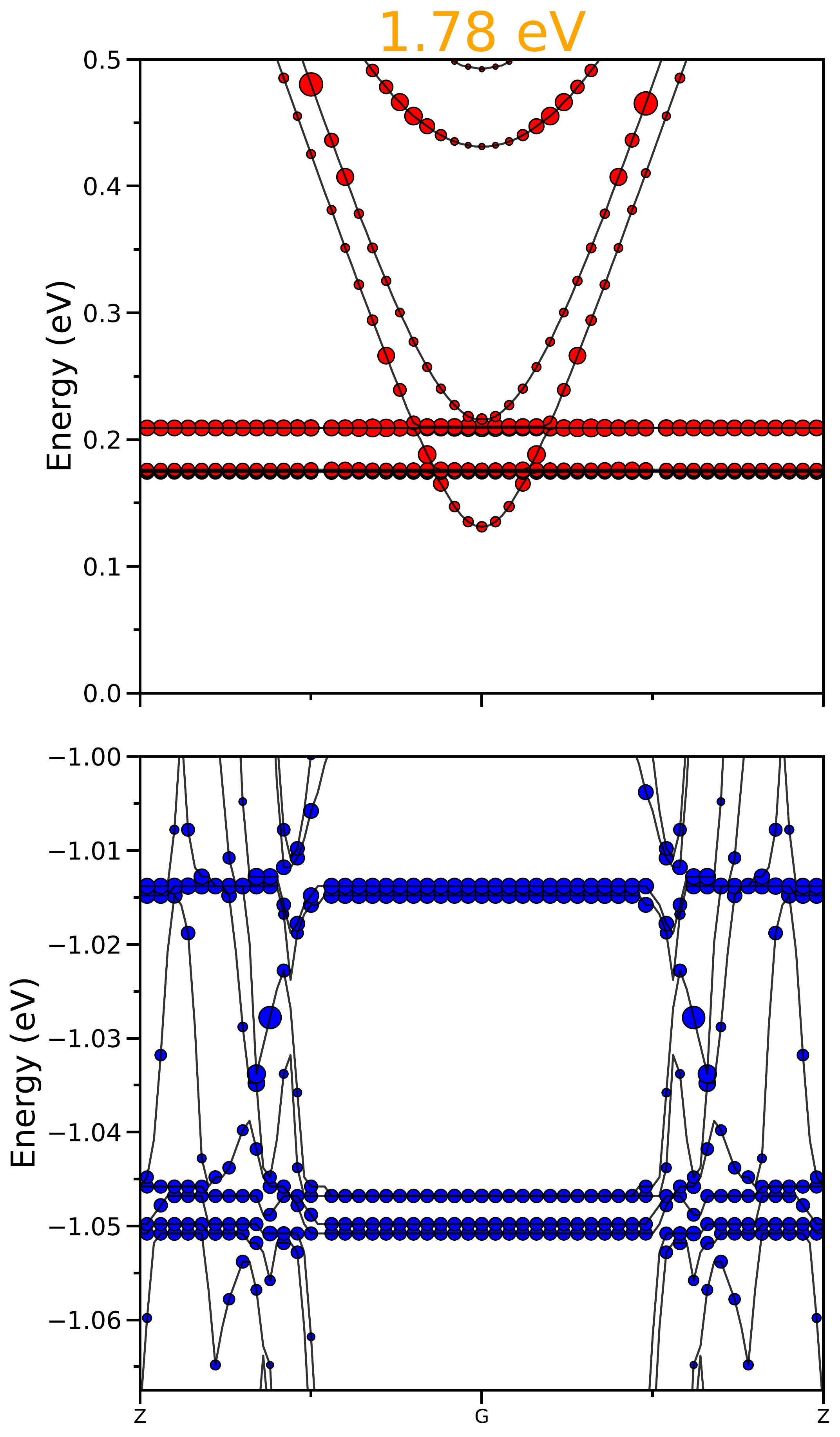}
    % \caption{Caption}
    % \label{fig:S_5}
\end{figure}
{{\bf Fig. S5. Detail of Fig. 2c} showing the band structure and populations for excitation at 1.78 in the energy regions corresponding to the HOMO and LUMO orbitals of the monolayer. Red (blue) circles denote the increase (decrease) of electron occupation and the circle size is proportional to the change of population.}

\newpage

\begin{figure}[hb!]
    \centering
    \includegraphics[width=.7\linewidth]{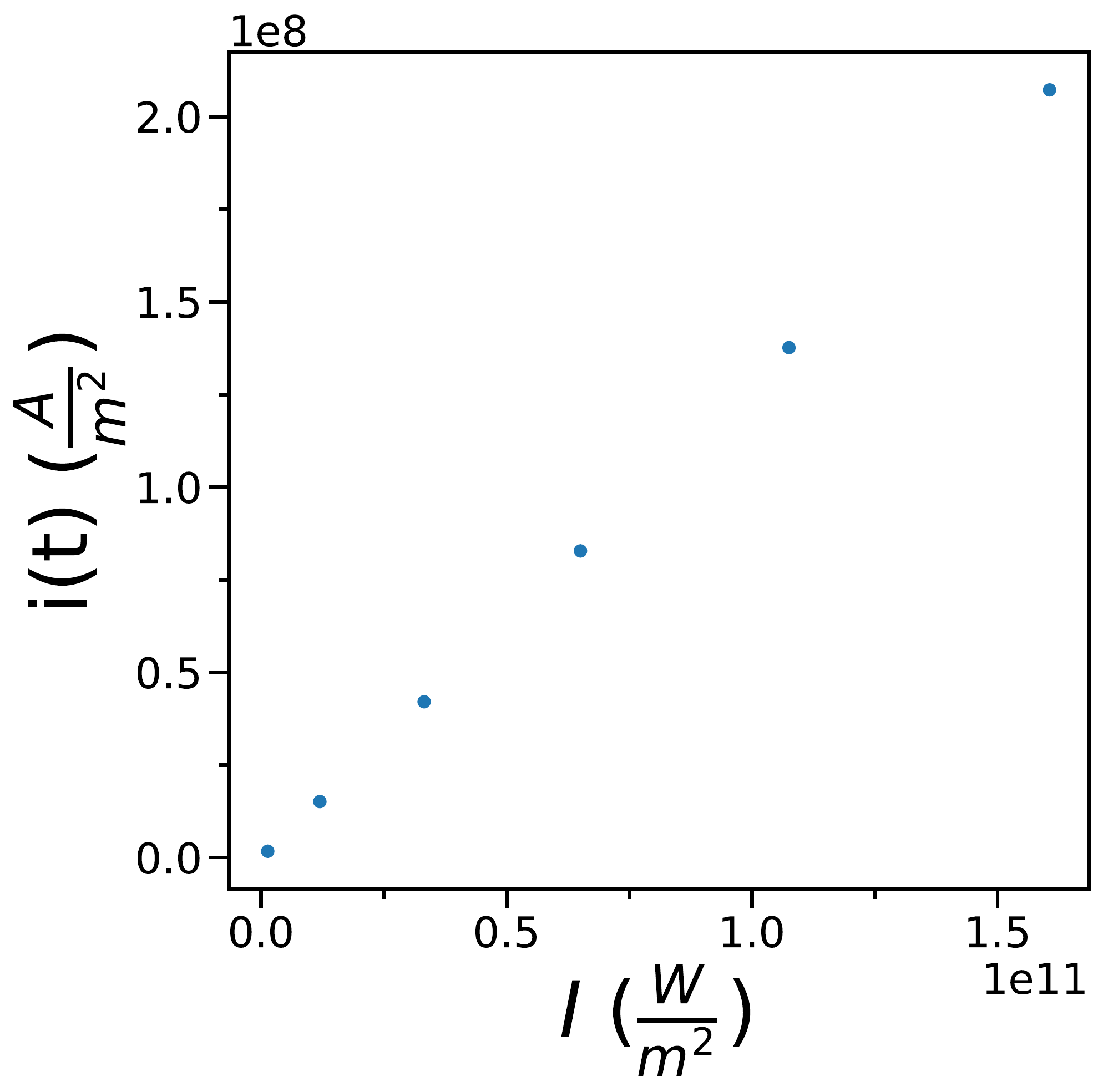}
    % \caption{Caption}
    % \label{fig:S_5}
\end{figure}
{\bf Fig. S6. Photoresponse shown as interlayer current as a function of laser intensity.} Currents where obtained from the stationary state (linear) portion of the time dependent charge shown in Figure S4.

\newpage

\bibliography{supplement}